\documentclass[%
 reprint,
 superscriptaddress,
 amsmath,
 amssymb,
 aps,
 longbibliography
]{revtex4-2}

\usepackage{subcaption}
\usepackage{graphicx}
\usepackage{dcolumn}
\usepackage{bm}
\usepackage{tikz}
\usepackage{braket}
\usepackage{amsmath}
\usepackage{appendix}
\usepackage{xcolor}
\usepackage{tikz}
\usetikzlibrary{quantikz}
\usepackage{hyperref}
\hypersetup{
    colorlinks=true,
    linkcolor=blue,
    filecolor=blue,   
    urlcolor=blue,
}

\usepackage{blochsphere}
\usepackage{algorithm2e}

\date{\today}

\begin{document}

\preprint{APS/123-QED}

\title{Qudits for Witnessing Quantum Gravity Induced Entanglement of Masses Under Decoherence}

\author{Jules Tilly}
 \email{jules.tilly.18@ucl.ac.uk}
 \affiliation{Department of Physics and Astronomy, University College London, WC1E 6BT London, United  Kingdom}
\author{Ryan J. Marshman}
 \affiliation{Department of Physics and Astronomy, University College London, WC1E 6BT London, United  Kingdom}
 
\author{Anupam Mazumdar}
 \affiliation{Van Swinderen Institute, University of Groningen, 9747 AG Groningen, The Netherlands}
\author{Sougato Bose}
 \affiliation{Department of Physics and Astronomy, University College London, WC1E 6BT London, United  Kingdom}

\begin{abstract}
Recently a theoretical and an experimental protocol known as quantum gravity induced entanglement of masses (QGEM) has been proposed to test the quantum nature of gravity using two mesoscopic masses each placed in a superposition of two locations. If, after eliminating all non-gravitational interactions between them, the particles become entangled, one can conclude that the gravitational potential is induced via a quantum mediator, i.e. a virtual graviton. In this paper, we examine a range of different experimental set-ups, considering different geometries and the number of spatially superposed states taken, in order to determine which would generate entanglement faster. We conclude that without decoherence, and given a maximum distance $\Delta x$ between any two spatial states of a superposition, a set of two qubits placed in spatial superposition parallel to one another will outperform all other models given realistic experimental parameters. Furthermore, when a sufficiently high decoherence rate is introduced, multi-component superpositions can outperform the two-qubit set-up. This is further verified with an experimental simulation, showing that $O(10^3)$ measurements are required to reject the no entanglement hypothesis with a parallel qubits set-up without decoherence at a 99.9$\%$ confidence level. The number of measurements increases when decoherence is introduced. When the decoherence rate reaches $0.125$~Hz, 6-dimensional qudits are required as the two-qubit system entanglement cannot be witnessed anymore. However, in this case, $O(10^6)$ measurements will be required. One can group the witness operators to measure in order to reduce the number of measurements (up to ten-fold). However, this may be challenging to implement experimentally.   

\end{abstract}

\maketitle


\section{Introduction}

Testing the quantum aspect of gravity is a central question of modern physics. While many theories of quantum gravity have been developed, there remains no consensus on how to unify the theories of general relativity and quantum physics. However, the lack of experimental evidence to gravity being quantum still remains an impediment in ongoing research \cite{Kiefer2012}. \\
\indent A number of recent experimental proposals have focused on trying to unveil General-relativity (GR) and post GR evidence \cite{Colella:1975dq,Nesvizhevsky:2003at,Stickler2018, Pikovski2012, Krisnanda2019, Howl2019, Carlesso2019, Altamirano2018,Marshman:2018upe}, and there has also been an initial attempt to rule out the semi-classical treatment of quantum gravity in~\cite{Page:1981aj}. Even attempts of detecting the B-mode polarisation of stochastic gravitational waves have uncertainties in the initial conditions for the Universe, which does not provide a concrete test for quantum nature of gravity~\cite{Ashoorioon:2012kh}.

In this regard, there has been recent progress in providing a razor-sharp witness to test the existence of the quantum nature of a graviton in a table-top experiment based on the observation that~\cite{marshman2020locality}:
\begin{itemize}
    \item The mediator of the universal gravitational interaction occurs via a spin-2 massless graviton, and if the graviton is quantum, it will entangle the two or more quantum matter states, and provide the static gravitational potential at the lowest order in the graviton/matter loop expansion. 
    \item The above statement strictly relies on two main assumptions: special relativity and perturbative quantum field theory, which allows an off-shell/virtual exchange of a graviton to mediate the gravitational force. 
    
\end{itemize}
\indent Based on this fact, a bonafide test for quantum nature of the gravitational interaction has been proposed in \cite{Bose2017}, where the two mesoscopic masses were allowed to interact in a spatially superposed quantum state via gravity. A similar proposal has also been made in \cite{Marletto2017}. This has attracted significant interest from the research community \cite{Belenchia2018, Christodoulou_2020, howl2020testing, Chevalier:2020uvv, Nguyen2019, Kamp2020,Chevalier:2020uvv, torovs2020relative,Miki:2020hvg,Matsumura:2020law,Rijavec:2020qxd,Toros:2020krn}, and an experimental initiative in creating macroscopic superposition with Stern-Gerlach set-up ~\cite{Margalit:2020qcy}. The above proposal has been coined as  Quantum Gravity induced Entanglement of Masses (QGEM), which exploits the loophole that as local operations and classical communications are unable to entangle the two quantum states if they were not entangled, to begin with, quantum communication is required to generate the entanglement as highlighted in \cite{marshman2020locality}. How a non-local gravitational interaction~\cite{Biswas:2011ar,Biswas:2005qr} entangles the two quantum states of matter is also shown in \cite{marshman2020locality}.

\indent This paper aims at analysing different possible set-ups for the QGEM proposal in order to determine which will be most efficient to implement in a real experiment. In particular, we consider how quickly different set-ups can generate entanglement according to a generalised model of the QGEM experiment and how many measurements would be required to witness that entanglement. In addition, proposals surrounding the QGEM experiment have been limited to qubits, so far. In this paper, we assess the implications of using quantum objects of higher dimensions (three: qutrits, $D$: qudits). We test our findings in the presence of decoherence, furthering the analyses presented in \cite{Nguyen2019}, \cite{Chevalier:2020uvv}, \cite{Kamp2020} and \cite{torovs2020relative}. \\
\indent Our key findings are that: the parallel set-ups \cite{Nguyen2019} of the experiment entangle faster than any other set-up considered; in the presence of decoherence, using qudits may be beneficial for the experiment, and could even be necessary; We provide an order of magnitude for the number of measurements required to reach a 99.9\% level of confidence (about 3.4$\sigma$) for different decoherence rates (up to $0.125Hz$). \\
\indent This paper will proceed as follows: Section \ref{sec:entanglement} presents a generalised version of the QGEM experiment, with arbitrary geometries and allowing for the use of spatial qudits. In Section \ref{sec:VNE} we analyse the different set-ups proposed using the entanglement entropy under the assumption of no decoherence. More practical entanglement witnesses are introduced in Section \ref{sec:EW} and used to re-analyse the experiment allowing for decoherence in Section \ref{sec:decoherence}. Finally, a statistical simulation of the results is presented in Section \ref{sec:expsim} to demonstrate how the required number of runs varies with the dimension of the qudits, and the decoherence rate assumed.


\section{QGEM with Qudits \label{sec:entanglement}}

	\begin{figure*}
	    \begin{subfigure}{.49\textwidth}
    		\includegraphics[width=1.0\columnwidth]{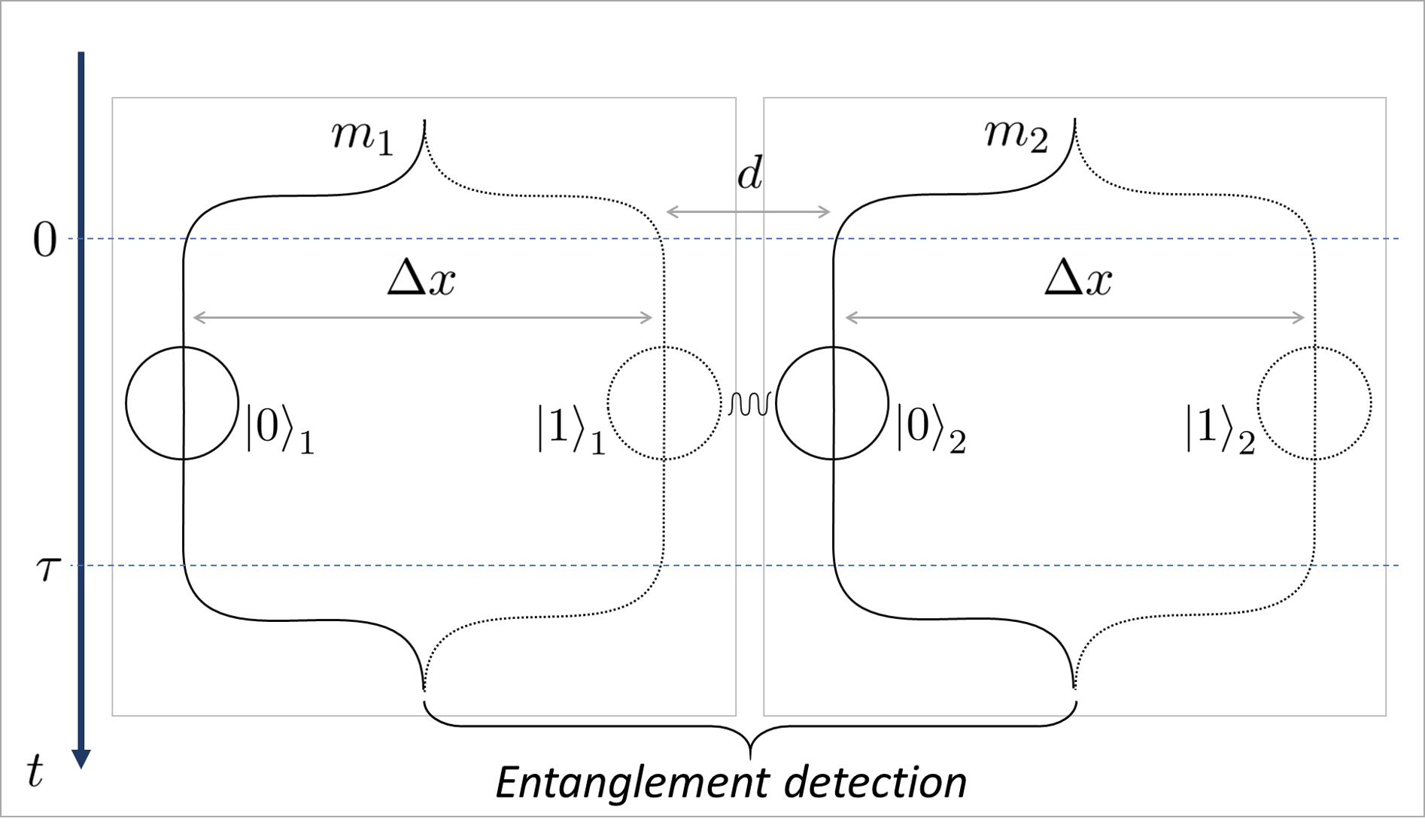}\par\medskip
    		\textit{Linear set-up.\label{fig:QGEMlinear1}}\par\medskip
		\end{subfigure}
	    \begin{subfigure}{.49\textwidth}
    		\includegraphics[width=1.\columnwidth]{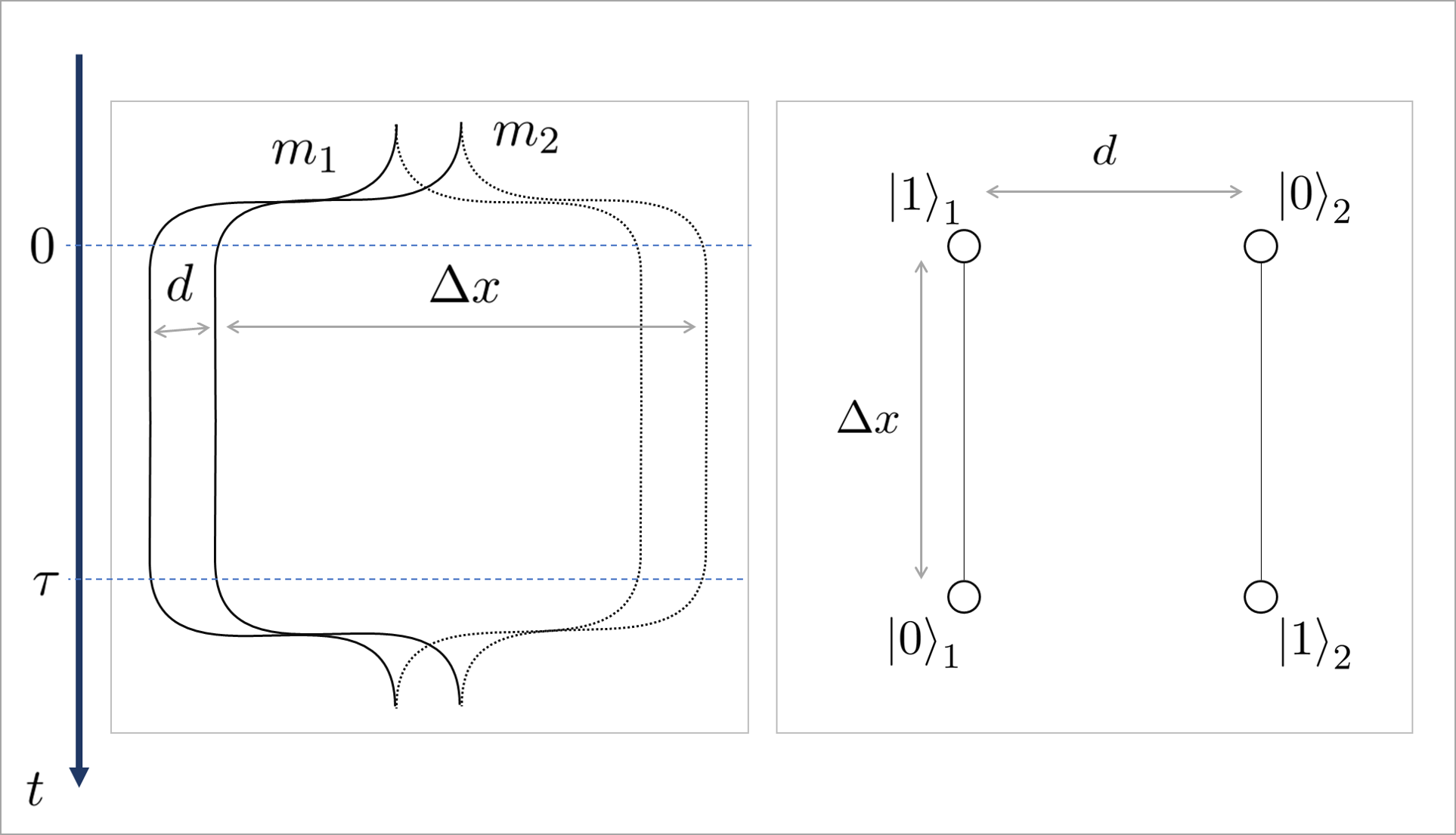}\par\medskip
    		\textit{Parallel set-up. \label{fig:QGEMpara1}}\par\medskip
		\end{subfigure}
		\caption{Schematic representation of the experiment, shown here using qubits. On the left, the liner set-up presented in \cite{Bose2017, Marletto2017, Kamp2020}. On the right, a schematic for the parallel set-up initially presented in \cite{Nguyen2019}}\par\medskip \label{fig:QGEMinit}
	\end{figure*}

\begin{figure}
	\centering
	    \par\medskip
		\includegraphics[width=1.0\columnwidth]{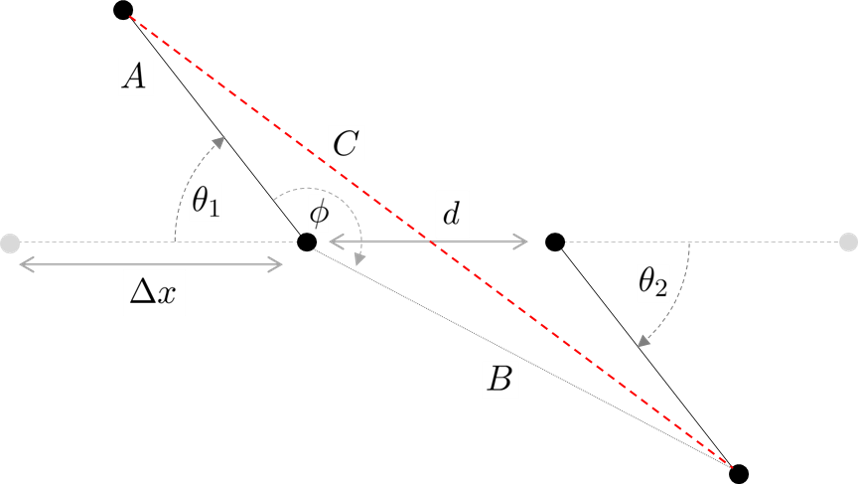}\par\medskip
		\caption{Top view for generalised QGEM model showing the maximum spread of the qudits (Distance $C_{pq}$ in red). The parametrization of the set-up is controlled by $\theta_1$ and $\theta_2$. \label{fig:generalised set-up}}
	\end{figure}

The QGEM experimental protocol~\cite{Bose2017,Kamp2020} ensured that the gravitational interaction would dominate over the electromagnetic interactions and the Casimir induced vacuum fluctuations. 
This also provides one of the primary constraints on the experimental set-up, which we will not seek to modify in this paper. Specifically, there must be some minimum distance $d$ maintained between the two particles. A schematic of two potential forms of the experiment is presented in Figure \ref{fig:QGEMinit}. \\
\indent The superposition width (or distance between the left-most and right-most superposition instance of each qubit) is labelled $\Delta x$, and was originally suggested to be $\sim 250 \mu m$\cite{Bose2017}. While the superposition is held, the qubits are maintained at a distance such that their inner-most superposition instances are a least $d$ apart (with $d\sim 200 \mu m$ \cite{Bose2017}). If $d<200 \mu m$ then forces such as Casimir-Polder forces and van der Waals forces can affect the overall state of the system: gravity is no longer the only possible quantum mediator for interactions between the two objects. There has been further work considered to mediate this (for instance, see: \cite{Kamp2020}), however, as it is beyond the primary considerations made here, we will not look to include this here as it does not affect any final conclusions. \\
\indent The two qubits are held in this superposition state for a time $\tau$ (\cite{Bose2017} suggests $\tau=2.5s$) after which the spatial states are brought together. The qubits are then measured to determine whether they were entangled by their gravitational interactions during the superposition period. \\
\indent Nguyen and Bernards proposed a nearly identical scheme \cite{Nguyen2019}, in which the superposition positions of each qubit are aligned parallel to each other as opposed to linearly in \cite{Bose2017,Kamp2020}. A schematic for this can be found in Figure \ref{fig:QGEMinit}. This scheme was motivated by the fact that maintaining the distance between the two qubits would be easier in the parallel case than in the linear case. \\
\indent In the remainder of this paper, we take the experimental parameters to match those proposed in Bose et al., that is, $d$ is always set to $\sim200 \mu m $ and the masses of the two qudits are always $\sim10^{-14} kg$. Furthermore, unless otherwise stated $\Delta x\sim250 \mu m$.\\ 
\indent While both~\cite{Bose2017} and \cite{Nguyen2019} have discussed the implementation of the respective set-ups, none of the intermediary models has been considered, and no direct comparison has been drawn with respect to their impact on how fast the qubit pair entangles or whether qudits would result in faster entanglement. \\
\indent We present a generalised model for the QGEM experiment, considering rotations of each interferometer, centred on the innermost spatial states of the two qudits (see Figure \ref{fig:generalised set-up}). The allowed space of the rotation angles, $\theta_1$ and $\theta_2$ must be restricted so that at no point the two masses come within a distance $d$ of one another. The linear set-up and the parallel set-up are special cases of the above, using the angles $\theta_1=\theta_2=0$ and $\theta_1 = \frac{3\pi}{2}, \theta_2 = \frac{\pi}{2}$ respectively. \\
\indent Our model also allows for the use of qudits, rather than qubits. That is, using $D$ spatial superposition states (and equivalently spin states, assuming Stern-Gerlach interferometry is used as previously proposed) where $D\ge2|D\in\mathbb{N}$. Under these conditions, the generalised state of the system resulting from the QGEM experiment can be written as:
		\begin{equation}
			\Ket{\psi(t = \tau)}=\frac{1}{D}\sum_{p = 0}^{D - 1}(\Ket{p}\otimes\sum_{q = 0}^{D - 1} e^{i\phi_{pq}} \Ket{q}).
		\end{equation}
Where: $\phi_{pq}$ is defined by:
		\begin{equation}
			\phi_{pq}\sim\frac{Gm_1m_2\tau}{\hbar C_{pq}}.
		\end{equation}
\indent The value $C_{pq}$ for each superposition pair can be derived using simple trigonometric rules and is given by

            \begin{equation}
			    C_{pq} = \sqrt{A_p^2 + B_q^2 - 2A_pB_q\cos(\theta_3)},
		    \end{equation}	
		    where
			\begin{equation}
			    A_p = ((D-1) - p)\frac{\Delta x}{(D - 1)},
		    \end{equation}
			\begin{align}
			    B_q &= [d^2 + \bigg(q\frac{\Delta x}{(D - 1)}\bigg)^2 \nonumber \\
			    &- 2d\bigg(q\frac{\Delta x}{(D - 1)}\bigg)\cos{(\pi -\theta_2)}]^{\frac{1}{2}},
		    \end{align}
			\begin{equation}
			    \theta_3 = \pi - \theta_1 +\arcsin{\bigg(\frac{q\frac{\Delta x}{(D - 1)}\sin(\theta_2)}{B_q}\bigg)}.
		    \end{equation}

\section{Entanglement entropy test \label{sec:VNE}}

\indent To assess different set-ups for the QGEM experiment, we begin by comparing the von Neumann entropy (VNE), or entanglement entropy, of the output state. We recall that noting $\rho_1$ the partial trace over the first qudit of the two-qudit system $\rho$, the entanglement entropy is given by \cite{Bengtsson_book}:
		\begin{equation}
			S(\rho) = -Tr[\rho_1log_2\rho_1],
		\end{equation}
or, using the eigen-decomposition of $\rho_1$: $\rho_1=\sum_{j}\lambda_j \Ket{j}\Bra{j}$ we can re-write $S(\rho)$ as:
		\begin{equation} \label{VNEformula}
			S(\rho) = -\sum_{j}\lambda_j log_2(\lambda_j).
		\end{equation}
\indent For a $D$ level system, the entanglement entropy is bound by:

		\begin{equation} \label{VQE_bounds}
			0\leq S(\rho) \leq log_2 (D),
		\end{equation}
		
From which it follows that more entanglement can be produced from higher dimension systems. However, this does not provide any information on how quickly such entanglement is created. Here we focus on the linear and parallel set-ups, a discussion regarding all other alternatives is presented in Appendix \ref{sec:discussgeneral}. \\
	\begin{figure*}
	    \begin{subfigure}{.49\textwidth}
	        \textit{Parallel set-up}\par\medskip
    		\includegraphics[width=1.\columnwidth]{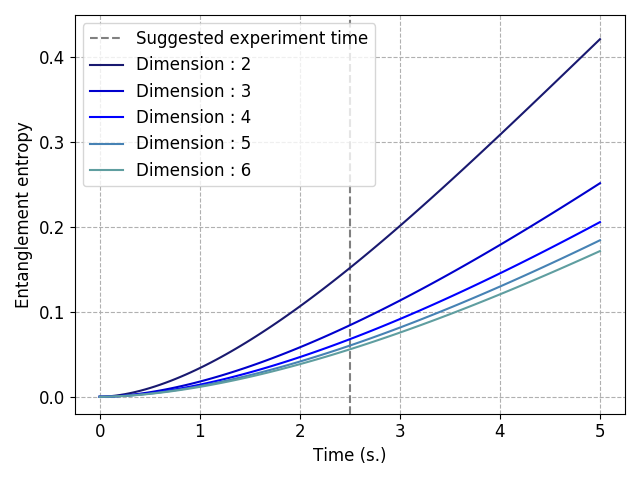}\par\medskip
		\end{subfigure}
	    \begin{subfigure}{.49\textwidth}
	        \textit{Linear set-up}\par\medskip
    		\includegraphics[width=1.\columnwidth]{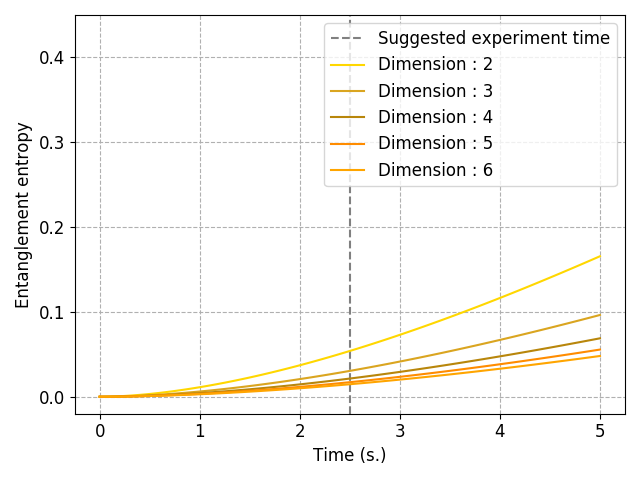}\par\medskip
		\end{subfigure}
		\caption{Entanglement entropy scalling with the state dimension for $D=2$ to $D=6$. On the left, we present the results for the parallel set-up, on the right we present the results for the linear set-up.} \label{para_qudits}
	\end{figure*}
\indent In order to assess whether qubits perform better or worse than higher dimensional qudits, we compute the entanglement entropy for $\rho_{para}$ and $\rho_{lin}$ using the generalised version of the model for 2 to 6 dimensional qudits. The resulting entropy scaling with time is plotted in Figure \ref{para_qudits}. The parallel set-up appears to perform significantly better than the linear set-up in realistic experiment times. \\
\indent At the proposed experimental time of $2.5$s, the qubit parallel case achieves a entanglement entropy of $0.152$, much larger than that for the qutrit case of $0.084$. Going to higher dimensions further reduces the entanglement entropy to $0.068$, $0.060$ and $0.056$ for 4, 5 and 6 dimensions respectively\footnote{Note that, at $\tau=2.5s$, as $D \to \infty$ the entanglement entropy converges to $\sim0.039$} suggesting that, under this model, multi-component superpositions do not entangle faster through gravity. \\
\indent Of course, given the superposition phases for the state resulting from the QGEM experiment are periodic, the linear set-up will achieve a higher entanglement entropy than the parallel set-up for sufficiently large experiment times. \\
\indent The entanglement entropy is not a valid metric for entanglement if classical mixing or decoherence are affecting the system as these are indistinguishable from entanglement as a source of entanglement entropy. Therefore, entanglement entropy is only used when assuming the system is at all times in a pure state. As such, for more realistic experiments, it is necessary to consider other methods of witnessing the entanglement.

\section{Entanglement Witness tests\label{sec:EW}}

\indent Multiple external factors can affect the system state throughout the experiment, these include decoherence and classical uncertainties introduced by the hardware used to implement the experiment.  Entanglement Witnesses provide a convenient testing system in the context of an experiment. \\
\indent The Positive Partial Transpose (PPT) entanglement witness is an appropriate witness for two for Negative Partial Transpose (NPT) entangled qudits:

		\begin{equation}
			\mathcal{W}_{ppt}=\Ket{\lambda_-}\Bra{\lambda_-}^T
		\end{equation}

\begin{figure}
        \includegraphics[width=1.\columnwidth]{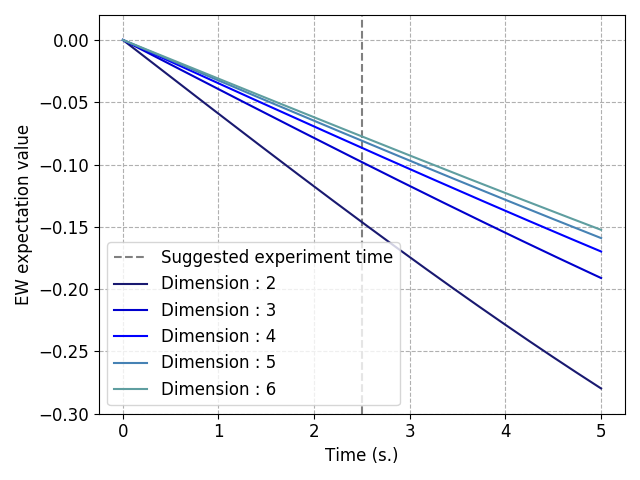}\par\medskip
		\caption{Expectation value of PPT entanglement witness for the parallel set-up with state dimension for $D=2$ to $D=6$}\par\medskip \label{fig:pptEW_qudit_para}
	\end{figure}

In the parallel qubit case, this witness is simply \cite{Chevalier:2020uvv}:
		\begin{equation}
			\mathcal{W}_{ppt} = \frac{1}{4}[\mathbb{I} - X\otimes X - Z \otimes Y - Y \otimes Z]
		\end{equation} 
\indent It is worth noting however that although this witness is not optimal in the linear set-up, having fewer terms to measure results in less variance when conducting the experiment (for a given number of measurements) (see Section \ref{sec:expsim}). Besides, all of the operators in the witness commute, as a result, there exist a measurement basis in which one can derive the expectation value of all these operators, reducing the number of terms to measure to one. \\
\indent When considering the qudit case, there exist states which are entangled but can not be detected by a PPT witness (see for example \cite{Horodecki1997, Horodecki1999}). These states have also been referred to as bound entanglement, and multiple methods have been developed to assess them~\cite{Horodecki1998, Yu2005,Guhne2006, Chruscinski2007, Chruscinski2011, Rudolph2003}. The witnesses described above may therefore not be sufficient to test entanglement of $D\otimes D$ systems (with $D>2$) as they would, by construction, fail to detect Positive Partial Transpose entangled states (PPTES). These can nonetheless be useful in case the theoretical density matrix of the state created by the experiment is indeed NPT. \\
\indent There is currently no general witness construction strategy for detecting PPTES (also known as entangled bound states) and research is focused on designing witnesses specific to certain families of states. Thankfully bound entanglement represents only a small proportion of all entangled states, and remains unlikely to occur in the proposed experiment. \\
\indent We computed the PPT entanglement witness expectation value for $D$ dimensional qudits with $D=2$ to $6$. The results are presented in Figure \ref{fig:pptEW_qudit_para} for both the parallel and linear cases.\\
\indent The expectation value of the PPT entanglement witness is nearing $-0.148$ at $\tau=2.5s$ for qubits. The witness values are slightly higher for higher dimensional qudits, echoing the findings of the entanglement entropy test. These witnesses also appear to be finest for the set of states that are produced by each set-up. A witness $\mathcal{W}_A$ is set to be finer than another witness $\mathcal{W}_B$ if it detects as entangled all the states detected by $\mathcal{W}_B$ and at least one more (see Ref. \cite{Lewenstein2000}). In this case, the witnesses computed are clearly able to detect all entangled states produced (for any $\tau$) albeit with values very close to zero for low values of $\tau$. \\
\indent Following the above, we can also note that the high dimension states generated by the QGEM experiment set-ups considered must be NPT entangled states; otherwise, $\mathcal{W}_{ppt}$ would fail to detect them as entangled. \\
\indent A further consideration regarding the entanglement witness is how many operators it needs to be broken into to be measured experimentally. For this, we can consider the scaling of generalised Gell-Mann matrices, which is a generalisation of the Pauli basis to quantum states of dimensions higher than 2. There are $D^2$ element for a set of Gell-Mann matrices for a $D$ dimensional quantum state. Because we build a system composed of two quantum objects, we are therefore looking at a maximum total number of operators of $D^4$ for the entanglement witness. \\
\indent We found that the entanglement witness derived from the PPT principle has, in general, a slightly lower than that number (due to the weights of certain operators being negligible or equal to zero in the decomposition of the entanglement witness). \\
\indent We can further significantly improve on this number by grouping together the operators that can be jointly measured. In general, operators can be jointly measured if they can be diagonalised together in a specific Tensor Product Basis (TPB) (for an overview of the method: \cite{Gokhale2019}. Conveniently this is equivalent to saying that operators can be jointly measured if they commute. Commutation is not transitive, and as such, there may be many different solutions to grouping operators together. To find a good solution, we use the Largest Degree-first Colouring (LDFC) algorithm (for a good summary of the LDFC algorithm, use the Supplementary materials in \cite{Hamamura2020}), whereby group are composed starting from one of the operators which have the highest number of commuting operators. The results we obtained are summarised in Table \ref{tab:table1}. \\

\begin{table}
\begin{ruledtabular}
\begin{tabular}{ ccc }
  D & PPT witness & PPT witness (grouped) \\
  \hline
  2 & 4 & 1   \\
  3 & 77 & 14  \\
  4 & 244 & 28 \\  
  5 & 613 & 53 \\
  6 & 1272 & 94\\  
\end{tabular}
\end{ruledtabular}
\caption{\label{tab:table1} Number of operators in the generalised Pauli decomposition of the witnesses, that must be measured to estimate the expectation value of the entanglement witness in the parallel case for $D=2$ to $D=6$. The grouped column present the number of operator groups that can be jointly measured in a single Tensor Product Basis, obtained using the Largest Degree First Coloring (LDFC) algorithm. Given this algorithm is a heuristic, one could find a different set and number of groups.}
\end{table}
\indent One point to note is that measuring operators jointly requires finding and implementing a joint measurement basis. While straightforward in some cases, this could yield some significant complications in an actual experiment.\\

\section{Testing models with decoherence \label{sec:decoherence}}

\begin{figure}
	\centering
		\includegraphics[width=1.\columnwidth]{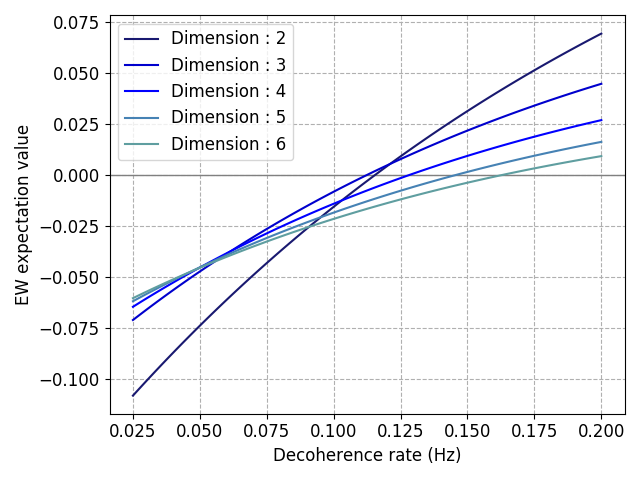}
	    \caption{Expectation value of PPT entanglement witness as a function of the decoherence rate in the parallel set-up and with $D$ ranging from 2 to 6.} \label{fig:qudit_deco}
		\par\medskip
	\end{figure}

\indent So far we have considered the case where both qubits are only coupled with each other through their positional superposition. Real experimental setting cannot however fully remove the potential for coupling of the studied quantum system with the environment.\\
\indent The particles' coherence and hence joint entanglement erodes over time due to interaction with the environment. This results in decoherence of the positional qudits into a single, defined position or a classical mixture of differing but well defined positions. \\
\indent We schematically incorporate this in our model by adding a time dependent exponential decay to all off diagonal term of each qudit's density matrix, parametrised by the decoherence rate $\gamma$. This is under the assumption that in any experiment $\frac{\Delta x}{D}\gg\lambda_{\textrm{dB}}$ where $\lambda_{\textrm{dB}}$ is the masses de Broglie wavelength. For a generic qudit, with dimension $D$ and density matrix $\rho_d$, we can write:
	\begin{equation}
	\resizebox{0.4\textwidth}{!}{$
        \rho_d = 
        \begin{bmatrix}
        c_{11} & c_{12} & \vdots & c_{1(d-1)} & c_{1d} \\
        c_{21} & c_{22} & \vdots & c_{2(d-1)} & c_{2d}\\
        \vdots & \vdots & \ddots & \vdots & \vdots\\
        c_{(d-1)1} & c_{(d-1)2} & \vdots & c_{(d-1)(d-1)} & c_{(d-1)d} \\
        c_{d1} & c_{d2} & \vdots & c_{d(d-1)} & c_{dd}
        \end{bmatrix}
        $}, \nonumber
	\end{equation}
as such, following the model described above, decoherence is incorporated as:
    \begin{equation}
    \resizebox{0.45\textwidth}{!}{$
        \rho_d \rightarrow 
        \begin{bmatrix}
        c_{11} & c_{12}e^{-\gamma \tau} & \vdots & c_{1(d-1)}e^{-\gamma \tau} & c_{1d}e^{-\gamma \tau} \\
        c_{21}e^{-\gamma \tau} & c_{22} & \vdots & c_{2(d-1)}e^{-\gamma \tau} & c_{2d}e^{-\gamma \tau}\\
        \vdots & \vdots & \ddots & \vdots & \vdots\\
        c_{(d-1)1}e^{-\gamma \tau} & c_{(d-1)2}e^{-\gamma \tau} & \vdots & c_{(d-1)(d-1)} & c_{(d-1)d}e^{-\gamma \tau} \\
        c_{d1}e^{-\gamma \tau} & c_{d2}e^{-\gamma \tau} & \vdots & c_{d(d-1)}e^{-\gamma \tau} & c_{dd}
        \end{bmatrix}
       $}. \nonumber
	\end{equation}
\indent The overall two-qudit system density matrix is then computed using: $\rho_{system}$ = $\rho_d^{(1)}\otimes\rho_d^{(2)}$.\\
\indent We computed the expectation value of $\mathcal{W}_{ppt}$ for dimensions 2 to 6 for incremental values of $\gamma$. The results are plotted in Figure \ref{fig:qudit_deco}.\\
\indent Interestingly, higher dimension models appear more resilient to decoherence than the qubit case. It is worth noting that increasing decoherence in the model also reduces the `optimal’ time for the experiment, that is, the time at which the entanglement witness is most negative, and the ability to detect entanglement with longer experiment times. To be better understand the interplay between time, decoherence and the number of dimensions, the expected value of $\mathcal{W}_{ppt}$ was computed at two different values for the decoherence rate: $0.1$Hz and $0.125$Hz.\\
	\begin{figure*}
	    \begin{subfigure}{.49\textwidth}
	    	\textit{$\gamma = 0.1$Hz}\par\medskip
    		\includegraphics[width=1.\columnwidth]{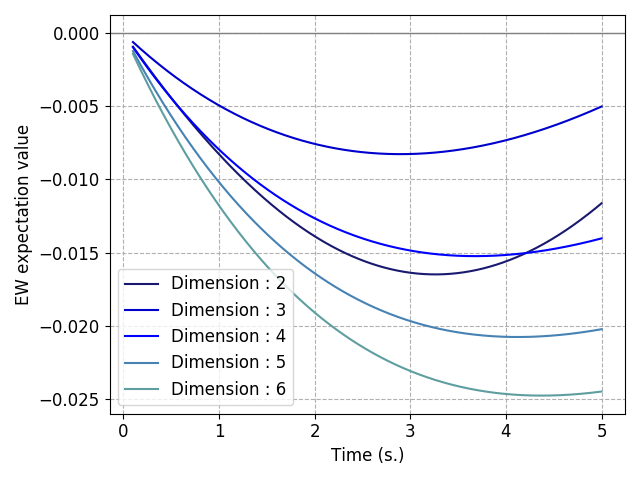}\par\medskip
		\end{subfigure}
	    \begin{subfigure}{.49\textwidth}
	    	\textit{$\gamma = 0.125$Hz}\par\medskip
    		\includegraphics[width=1.\columnwidth]{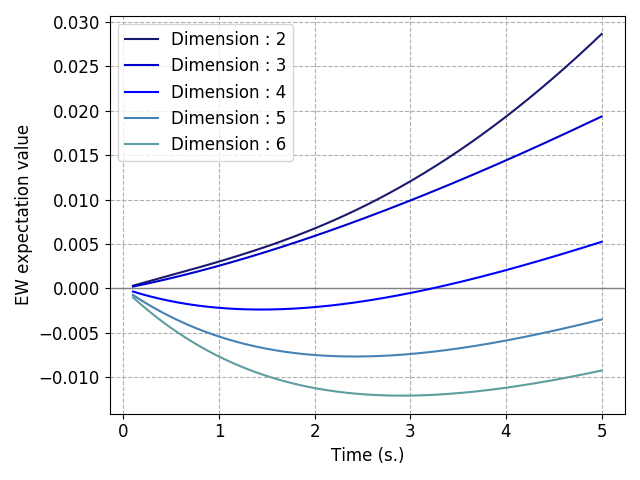}\par\medskip
		\end{subfigure}
	    \caption{Entanglement witness expectation value over time for the parallel set up and with $D$ ranging from 2 to 6. We present two different possible values of $\gamma$: 0.1Hz, and 0.125Hz}
	\end{figure*}
\indent The advantage of higher-dimensional models, when decoherence is increased, become significant when $\gamma\gtrsim0.1$Hz. This suggest that in a real run of the experiment, multi-component superposition may be preferable, or even necessary, if decoherence is sufficiently high. We can also observe as expected that longer time becomes detrimental for high decoherence rates. We also considered the problem of optimising the experiment run-time to maximise the decoherence rate for which the entangled witness could detect the state as entangled. In Appendix \ref{sec:time_opt} we plot the expectation value of $\mathcal{W}_{ppt}$ for different values of $\tau$, including decoherence. The result is that $\tau=2.5$s is near optimal for qubits and qudits, though lower time performs marginally better for the latter. Of course the experiment can be run for shorter times without any negative impacts provided a detectable level of entanglement has developed. \\
\indent While higher dimension set-ups are more resilient to decoherence, one key question in this analysis is to determine what number of measurements will be required to reject that the experiment state is not entangled. In order to further this analysis, we now need to fully simulate the experiment in order to determine the required number of measurements and run-time of the QGEM experiment. This is the object of the next section. 
\section{Experiment simulation\label{sec:expsim}}

\begin{figure}
		\includegraphics[width=1.\columnwidth]{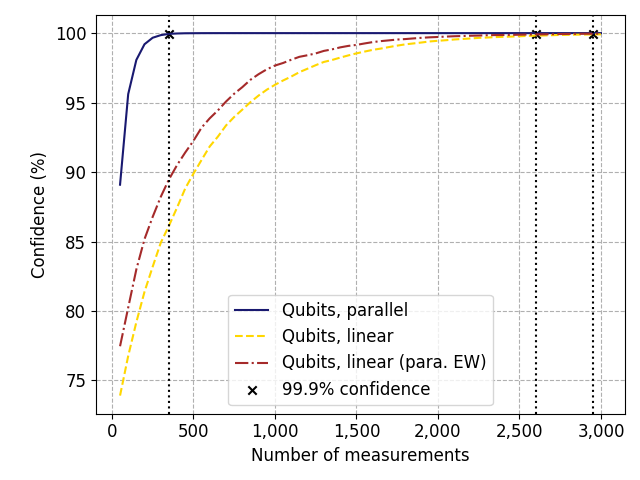}\par\medskip
	    \caption{QGEM experiment simulation - linear and parallel set-ups (no decoherence) - Expected level of confidence (probability of the witness value actually being negative) in the qubit case. We added the case in which the parallel witness is used for the linear set-up showcasing the benefits of having a lower number of terms to compute}\label{fig:expsimqubits}
	\end{figure}

\begin{figure}
    		\includegraphics[width=1.\columnwidth]{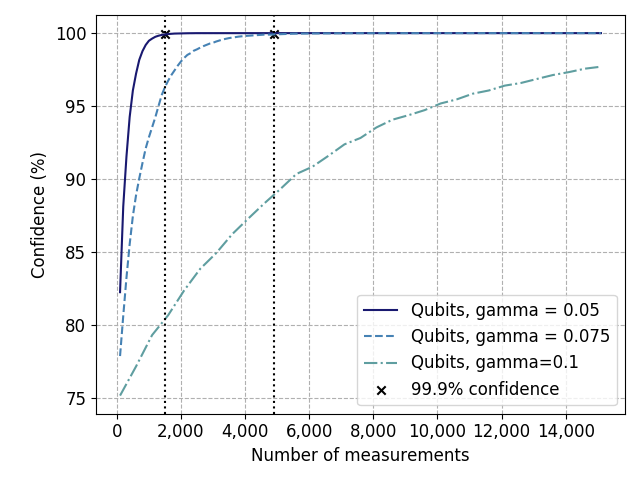}\par\medskip
	    \caption{QGEM Experiment simulation with qubits - Confidence levels for three decoherence level (0.05Hz, 0.075Hz and 0.1Hz) as a function of the number of measurements available. Here the witness as expectation value of -0.074, -0.043, and -0.016, respectively} \label{fig:expsim_qubits_deco}
	\end{figure}

\begin{figure}
    		\includegraphics[width=1.\columnwidth]{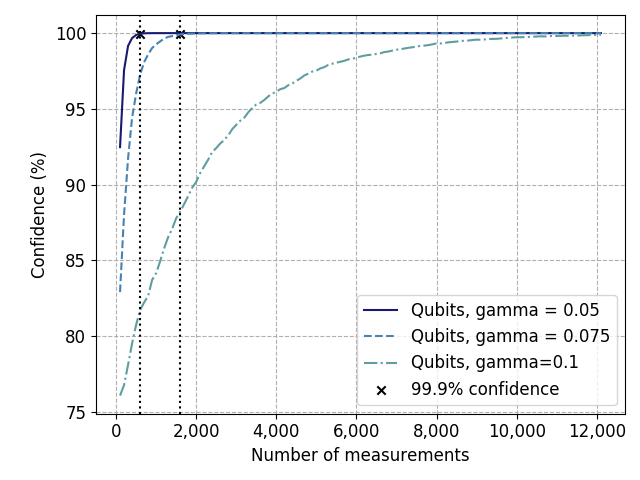}\par\medskip
	    \caption{QGEM Experiment simulation with qubits (operators grouped) - Confidence levels for three decoherence level (0.05Hz, 0.075Hz and 0.1Hz) as a function of the number of measurements available. Here the witness as expectation value of -0.074, -0.043, and -0.016, respectively} \label{fig:expsim_qubits_grouped}
	\end{figure}

\begin{figure}
    	\includegraphics[width=1.\columnwidth]{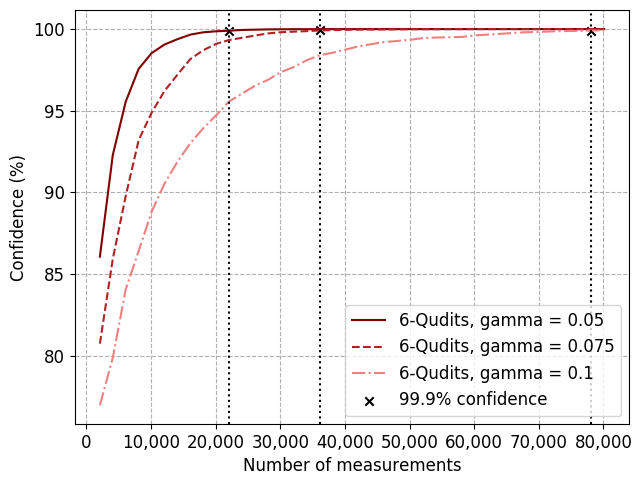}\par\medskip
	    \caption{QGEM Experiment simulation with 6-qudits (operators grouped) - Confidence levels for three decoherence level (0.05Hz, 0.075Hz and 0.1Hz) as a function of the number of measurements available. Here the witness as expectation value of -0.045, -0.032, and -0.021, respectively} \label{fig:expsim_6qudits_grouped}
	\end{figure}

\indent In this section, we present experiment simulations used to estimate the number of measurements that will be required to reject the hypothesis that the qubit pair is not entangled. Failure to reject only means that it is impossible at this stage to confidently test for gravitationally mediated entanglement. \\
\indent To compare the entanglement witnesses, and the different model proposed, we simulate an experiment, as described below: 
	\begin{enumerate}
		\item The entanglement witness is decomposed into a weighted sum of generalised Pauli tensors terms (Gell-Mann matrices for qutrits, and generalised $d$ dimension Pauli operators for qudits). (We used the method for generalisation of Pauli operators described in Thew et al. \cite{Thew2002}). In some cases, we group the operators (following the groups described in Table \ref{tab:table1}).
		\item The quantum state resulting from the experiment is measured a pre-determined number of times against each of the Pauli terms or group, to compute (1) their expectation value, (2) the standard error of the measurement series. To minimise the variance of the observable for a given number of measurements, we have weighted the number of measurement in proportion to the weight of each Pauli tensor terms (or group) in the decomposition of the witness.
		\item Details on how confidence levels are computed can be found in appendix \ref{sec:stats}.
	\end{enumerate}
\indent The plots presented in this section are an average over many single runs of each numerical experiment simulation. As such, these should be representative of a typical run of the experiment. The conclusions drawn from these are only meant to indicate the order of magnitude of the number of measurements required in order to define the most adequate experiment set-up. \\
\indent We first describe the result obtained from the experiment simulation with qubits. The simulation is first run comparing the linear and parallel set-ups with no decoherence (Figure \ref{fig:expsimqubits} followed by plots of the simulation for increasing decoherence rate $\gamma$  in Figure \ref{fig:expsim_qubits_deco}. We have used $\mathcal{W}_{ppt}$ with full basis decomposition in this first experiment simulations, holding $\Delta x = 250 \mu m$ and $\tau = 2.5s$.\\
\indent As we can see from Figure \ref{fig:expsimqubits}, with no decoherence, it takes about 500 measurements to reject the hypothesis that the two states are not entangled at a 99.9\% confidence level in the parallel case and at least 3,000 measurements for the linear set-up.  Using the witness derived in the parallel set-up marginally improves the results of the linear case, although not sufficient for it to be comparable to the parallel version of the experiment. This further confirms that the parallel set-up will be preferable in a real experiment and we, therefore, discard the linear set-up in the remainder of the simulations. \\
\indent We can expect that incorporating decoherence will increase the number of measurements required as it pushes the expectation value of $\mathcal{W}_{ppt}$ upward. The result of the qubit experiment simulations with decoherence are presented in Figure \ref{fig:expsim_qubits_deco}. It illustrates the rapid increase in the number of measurements required as the decoherence rate is raised. For $\gamma=0.05$Hz, the parallel set-up still only require about $\sim 2,000$ measurements. This figure goes up to $\sim 6,000$ measurements for $\gamma = 0.075$Hz. At $\gamma=0.1$Hz, the experiment would required at least $25,000$ measurements. At this decoherence rate, qubits have fewer negative expectation values for $\mathcal{W}_{ppt}$ than some qudits, and at decoherence rate above $\gamma = 0.12$Hz, the expectation value of the witness is positive (that is, no entanglement is detected and therefore the results are not plotted). \\
\indent We repeated the experiment simulations as described above with differing decoherence rates in the case of qudits. We have used the $D=6$ qudit case for illustration. For brevity, we will denote them as 6-qudits, and use an analogous nomenclature for other dimensions. As for the qubits simulations, all the experiments use the parallel set-up, $\Delta x = 250 \mu$m and $\tau = 2.5$s.\\
	\begin{figure}
    		\includegraphics[width=1.\columnwidth]{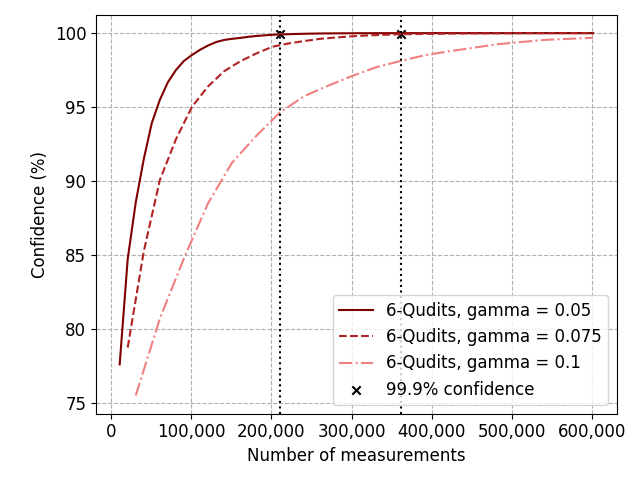}\par\medskip
	    \caption{QGEM Experiment simulation with D=6 qudits - Confidence levels for three decoherence level (0.05Hz, 0.075Hz and 0.1Hz) as a function of the number of measurements available. Here the witness as expectation value of -0.045, -0.032, and -0.021, respectively} \label{fig:expsim_qubits_deco_6d}
	\end{figure} 
\indent As expected, for the 6-qudits case, the number of measurements required is significantly higher. This is primarily due to the large number of additional terms to compute. Our results are presented in Figure \ref{fig:expsim_qubits_deco_6d} and show that for $\gamma = 0.05$Hz, over 200,000 measurements would be needed, while nearly 400,000 would be required if $\gamma = 0.075$Hz, and about 600,000 for $\gamma = 0.1$Hz. \\
\indent One last test that is worth considering is the situation in which grouping of terms is allowed (this is subject to being able to produce the relevant measurement bases experimentally, as mentioned in Section \ref{sec:EW}). \\
\indent Figure \ref{fig:expsim_qubits_grouped} presents the results for the qubits case. We can already see the drastic reduction in the total number of measurements required, resulting from the reduction of terms to measure from 4 (in reality 3 since the identity term does not need to be measured), to 1. Less than 1,000 measurements and 2,000 measurements are necessary when $\gamma = 0.05$Hz and $\gamma = 0.075$Hz, respectively. Similarly, only about 12,000 measurements is required if $\gamma = 0.1$Hz.\\
\indent A very similar pattern can be seen in the qudit case. The results for 6-qudits are presented in Figure \ref{fig:expsim_6qudits_grouped}. number of measurements are drastically reduced when operators are grouped. About 25,000 measurements are needed for $\gamma = 0.05$Hz, while $\gamma = 0.075$Hz requires less than 40,000 measurements, and $\gamma = 0.1$Hz less than 80,000.

\begin{figure}
    		\includegraphics[width=1.\columnwidth]{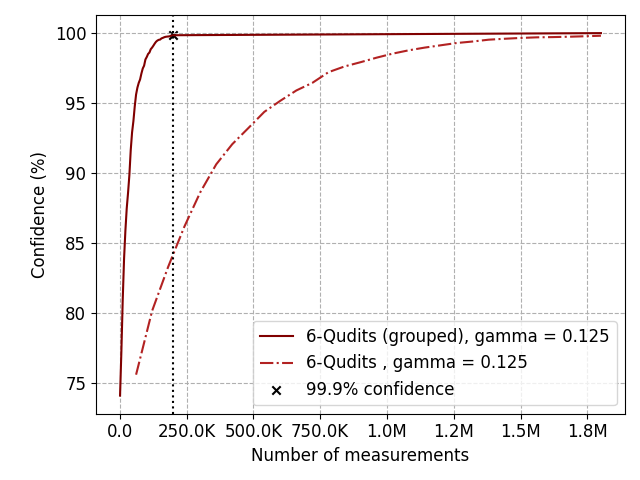}\par\medskip
	    \caption{QGEM Experiment simulation with 6-qudits - Confidence levels for a decoherence rate of 0.125Hz as a function of the number of measurements available in a case where the operators are grouped, and not.} \label{fig:expsim_qudits_125}
	\end{figure}

\indent There are clearly no reasons to use high dimensional qudits unless the decoherence rate is such that the entanglement witness for qubits always has a positive expectation value. In the experiment setting, we have assumed, this happens at about $\gamma \approx 0.12$Hz. We therefore present in Figure \ref{fig:expsim_qudits_125} an experiment simulation showing how 6-qudits perform in the window of decoherence rate in which high dimensional qudits become relevant.
\indent At $\gamma = 1.25$Hz, it is clear that the 6-qudits states still require a very large number of measurements in order to reject the null hypothesis. However, grouping the operators to measure in joint measurement bases allow reducing the number of measurements required from nearly 2,000,000 to slightly above 200,000. It is worth noting that at this rate, the qubits case would not work as decoherence pushes the expectation value of the witness above zero, and the 4-qudits and 5-qudits case are too close to zero to reach confidence of $99.9\%$ in a realistic number of measurements. \\

\section{Conclusion\label{sec:conclusion}}

\indent This paper considered modifying the set-up proposed in the original QGEM experiments to determine how best to generate, protect and detect entanglement in a future experiment. We looked at two aspects, in particular, the geometric set-up of the experiment, and the number of dimensions of the quantum objects used and developed a generalised mathematical model of the experiment. \\
\indent Based on this model, and using entanglement entropy, we concluded that the parallel qubits set-up generates entanglement the fastest for realistic experiment run-times ($\tau$ $O\left(1\mathrm{s}\right)$). As entanglement entropy cannot account for entanglement once decoherence is introduced, we presented an entanglement test based on entanglement witnesses. We concluded that as the decoherence rate is increased higher dimension, qudits finally out-perform qubits by providing lower expectation values for the witnesses. \\
\indent To estimate which set-up would require the least number of measurements to evidence entanglement, we simulated experiments to define a confidence level for the negativity of the expectation value of the witness given a certain number of measurements. Without noise, at a 99.9$\%$ certainty level, the parallel qubits set-up requires less than 2,000 measurements (1,000 when grouped) to reject the null hypothesis (that the state is not entangled). The number of measurements required increases rapidly when decoherence is introduced. \\
\indent Qudits of higher dimensions only become more useful than qubits when the expectation value of the witness for qubits becomes non-negative. That is because the number of basis elements to be estimated to calculate the witness expectation value increases quadratically in the number of dimensions ($d^4$). In the experiment settings proposed, at $\tau=2.5$s and $\Delta x = 250 \mu$m, qudits of dimension 6 are more favourable than qubits when the decoherence rate is $\gamma\sim0.125$Hz, however, in this case, over 2,000,000 measurements will be required (200,000 when operators are grouped). \\
\indent Thus to further improve the experiment design, the following points should be considered. Clearly, reducing the qudit-pair exposure to the environment to decoherence would render the experiment more economical. The total expected decoherence rates in specific, realistic experimental designs must be estimated in order to confirm whether higher dimension qudits will be required. Finally, any increase in the superposition width $\Delta x$ is significantly beneficial as it both improves the entanglement generation rate, allowing a significant reduction in overall run-time, which therefore also minimises the impact of decoherence. 

\section{Acknowledgements}
JT is supported by the UK EPSRC grount No. EP/R513143/1. RJM is supported by a University College London departmental studentship. AM’s research is funded by the Netherlands Organisation for Science and Research (NWO) grant number 680-91-119. SB would like to acknowledge EPSRC grants No. EP/N031105/1 and EP/S000267/1.

\newpage

\begin{appendices}
\begin{center}
    \textbf{APPENDICES}
\end{center}
	
\section{Discussion on the geometrically generalised model \label{sec:discussgeneral}}

\indent There are a number of parameters in the generalised formula that can be modified and tested for more effective entanglement generation. We have isolated two which offer particular insights, the superposition width ($\Delta x$) in Section \ref{sec:superposition_width} and the rotation angles ($\theta_1$ and $\theta_2$) in section \ref{sec:angles}. Prior to this, however, there are a few points that are worth noting:\\
\begin{enumerate}
    \item It is clear that more time can allow achieving maximally entangled states, however increasing the experimental run time is unrealistic due to greater risk of decoherence. This point is covered in more details in Section \ref{sec:decoherence}.
    \item Reducing the minimum distance $d$ also clearly generates much faster entanglement. However, as mentioned before this is not necessarily useful in practice. This has been considered by others~\cite{Kamp2020}, and any results here will also hold for such modified set-ups.
    \item Using more massive quantum objects result in higher relative phases and in faster entanglement generation. However, more massive objects would also mean more challenging implementation for the interferometry, larger particle radii and hence higher Casimir-Polder forces. This could, in turn, increase the minimum distance $d$ and overall negatively impact entanglement growth. This is not something we will consider further.
\end{enumerate}

\subsubsection{Comparing rotation angle set-ups\label{sec:angles}}

\begin{figure}
	\centering
	    \includegraphics[width=1.0\columnwidth]{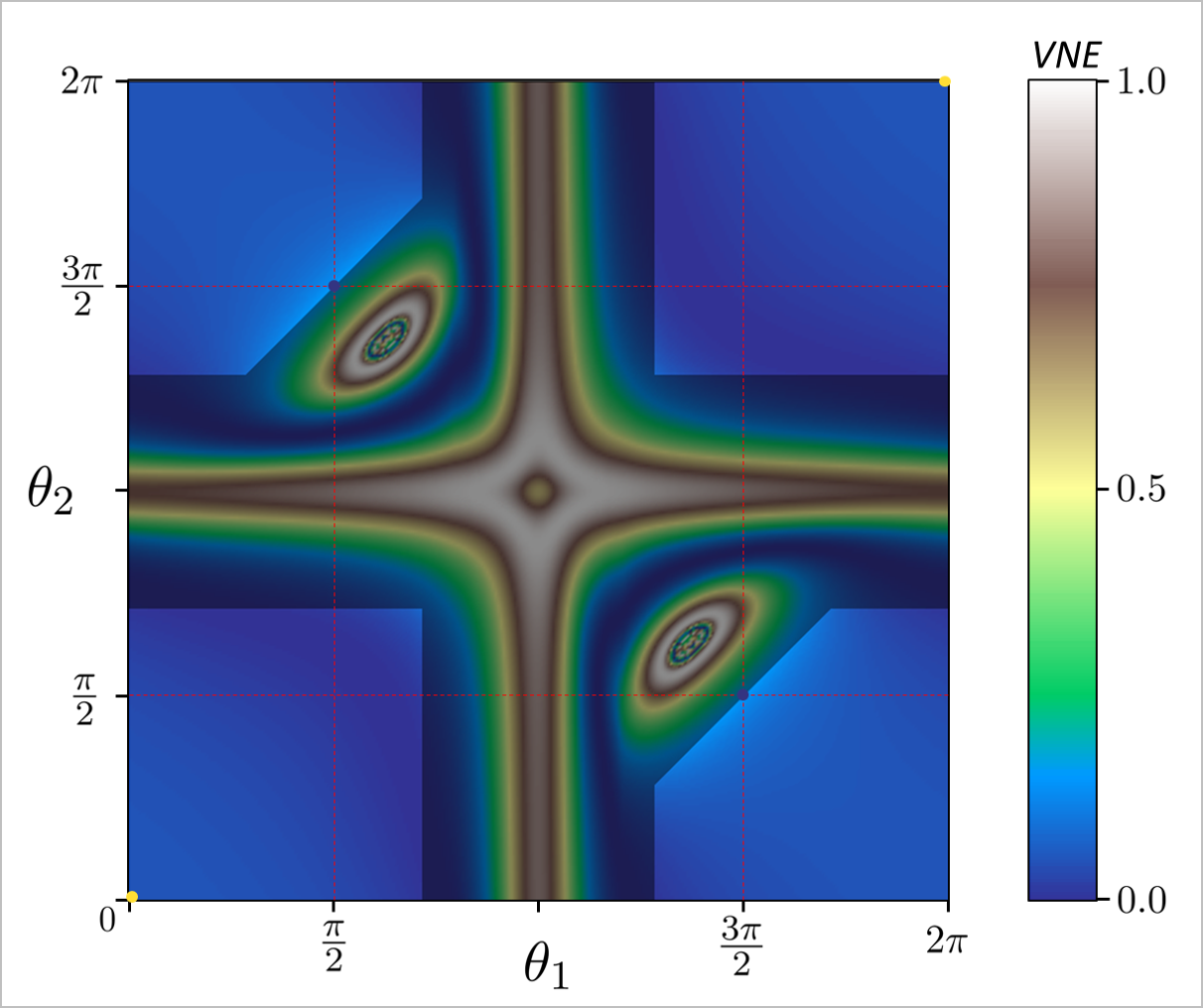}\par\medskip
	    \caption{Entanglement entropy for all possible combinations of $\theta_1$ and $\theta_2$. The areas of reduced contrast (the region around the center $+$) are forbidden based on the requirement that the states do not come too close.}\label{heatmap2}
	\end{figure}

\indent In the main body of the paper, we have only presented the linear and parallel cases. Based on our results, the parallel set-up entangles the qudits faster. For completeness, and to allow for further modification in the implementation by experimentalists, we also considered a range of additional geometries for the set-up that can be implemented using the generalised QGEM model derived previously. \\
\indent The heat-map presented in Figure \ref{heatmap2} displays the entanglement entropy for all possible combinations of $\theta_1$ and $\theta_2$ as defined in Figure \ref{fig:generalised set-up}, having set $\Delta x = 250\mu$m and $\tau = 2.5$s. The dark blue dot represents the two possible parallel set-ups ($\theta_1 = \frac{3\pi}{2}, \theta_2=\frac{\pi}{2}$ and $\theta_1 = \frac{\pi}{2}, \theta_2=\frac{3\pi}{2}$), while the yellow dot represents the linear set-up ($\theta_1 = \theta_2 = 0$). On the heat-map, blue represents low entanglement entropy while white represents entanglement entropy nearing 1.0. It appears from this figure that higher entanglement entropy could possibly be achieved with alternative set-ups; however, this is without considering that some combinations result in some superposition instances being under $200 \mu m$ and therefore subject to non-negligible Casimir-Polder forces (represented by the shaded areas on the figure).

\subsubsection{Impact of superposition width ($\Delta x$) on entanglement  entropy\label{sec:superposition_width}}

\begin{figure}
	\centering
	    	\includegraphics[width=1.0\columnwidth]{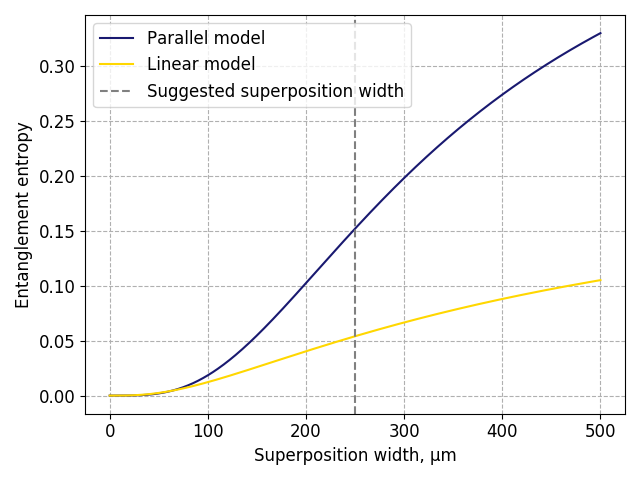}
	    	\caption{Entanglement entropy for the parallel and linear set-ups in the qubit case as a function of the superposition width ($\Delta x$)}\par\medskip \label{fig:superposition_width}
	\end{figure}

\indent Figure \ref{fig:superposition_width} shows that a larger superposition width results in faster entanglement entropy growth. To illustrate this point we can consider the phases for the qudit parallel set-up: $\phi_{pq}$ is smaller or equal to $\phi$, reaching equality at $\Delta x=0$m, as $\Delta x$ increases, $\phi_{pq}$ decreases resulting in all the phase factors $\Delta_{\phi_{pq}}$ to become more negative than they already are and as such accelerating entanglement generation. 

\begin{figure}
	\centering
		\includegraphics[width=1.0\columnwidth]{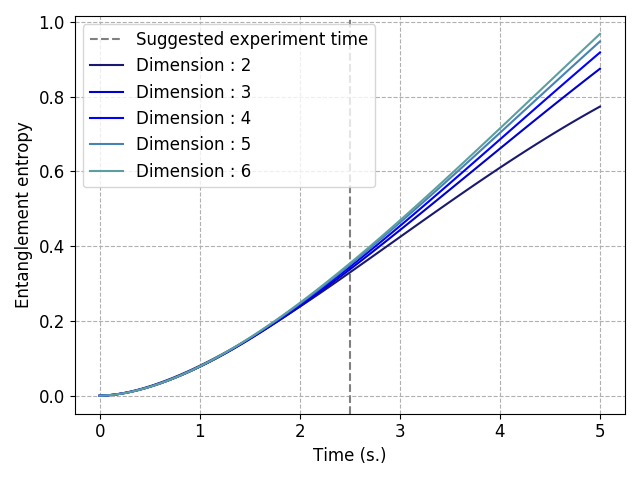}
	    \caption{Entanglement entropy as a function of time for $D$ ranging from 2 to 6. In this case, we have scaled $\Delta x$ to the number of dimensions} \label{fig:scaling}
	\end{figure}

\indent The superposition width, however, is limited in practice by experimental considerations such as the magnetic field gradient achievable. This suggests that by further modifying the arrangement into a parallel set-up one can further reduce the magnetic field gradient and/ or other experimental parameters while maintaining a detectable level of entanglement.\\
\indent Using higher dimension spin quantum objects would, in theory, result in larger possible superposition width due to the spin-dependent nature of the magnetic field gradient coupling. This suggests that higher-dimensional objects would perform better than qubits if their maximum superposition width is higher. However, once a larger spin object is created, one can simply initialise the state including only the outer spin states and hence recreating a qubit state which performs better at an equivalent value of $\Delta x$. For illustration purposes, Figure \ref{fig:scaling} represents a plot of entanglement entropy against time for qudits in which $\Delta x$ is scaled to the number of dimensions. This somewhat unrealistic set-up leads to higher-dimensional objects perform significantly better.

As such, the analysis allows to conclude that: 
	\begin{enumerate}
		\item Qubits entangle faster than qudits for values of $\tau$ and $\Delta x$ realistic for the experiment.
		\item Parallel set-ups entangle faster for all dimensions.
		\item The superposition width $\Delta x$ will be the variable of interest in terms of obtaining faster entanglement or a more easily implementable experiment.
	\end{enumerate}

\section{Alternative witness: Vicinity witness for Negative Partial Transpose entangled states} \hfill \break

\indent We consider an alternative entanglement witness, which is built to detect entangled states in the vicinity of a known entangled pure state. Constructing this witness amounts to finding a value for $\alpha$ such that $ Tr(\mathcal{W}_{vic}\varrho)\geqslant0$ for all separable states, with $\mathcal{W}_{vic}$ given by:
		\begin{equation}
			\mathcal{W}_{vic}=\alpha\mathbb{I} - \Ket{\psi}\Bra{\psi}
		\end{equation}

\indent The maximum value of $\alpha$ is then derived as the square of the maximum Schmidt coefficient of the pure state~\cite{Bourennane2004}. Noting $\lambda_m$ the highest of these coefficients, we can re-write the witness as:

		\begin{equation}
			\mathcal{W}_{vic}=\lambda_m^2\mathbb{I} - \Ket{\psi}\Bra{\psi}
		\end{equation}

\indent We can now compare the performance of the PPT based witness $\mathcal{W}_{ppt}$ and the `vicinity' witness $\mathcal{W}_{vic}$. This analysis is restricted to the qubit, parallel version of the QGEM experiment as it appears to be the most optimal set-up. 

	\begin{figure}
	\centering
	    \includegraphics[width=1.0\columnwidth]{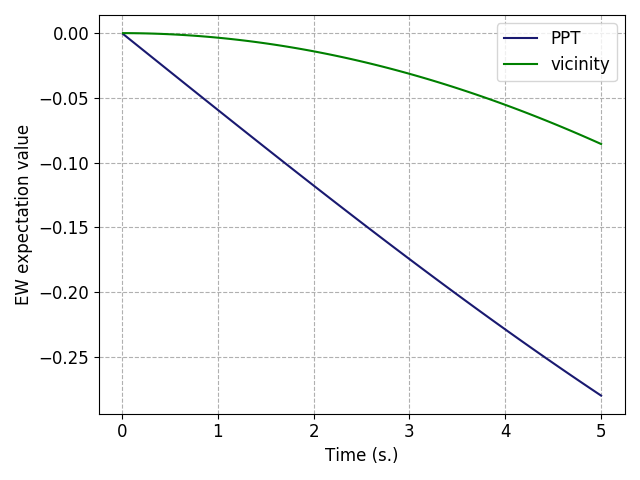}\par\medskip
	\caption{Expectation value of PPT and Vicinity entanglement witnesses as a function of time in the qubit case}\par\medskip \label{fig:EW_comp}
	\end{figure}

\indent Comparing the expectation value of the PPT-based witness and the vicinity witness, the former exhibits clearly much lower values on short time frames. As such, it is the recommended entanglement witness for the implemented experiment. In Table \ref{tab:table_2}, we show that the vicinity witness, in general, requires to measure fewer terms than the PPT witness - it is not sufficient however to make it more advantageous than the PPT witness in an experimental model, and therefore we have mostly not included it in our analysis. \\
\indent There also exist numerous other entanglement witnesses, often based on existing separability criteria, which will not be treated here as not directly relevant to the proposed research.\footnote{For an overview of several entanglement witnesses, Ref.~\cite{Guhne2009}}

\begin{table}
\begin{ruledtabular}
\begin{tabular}{ cccc }
  D & PPT (para.) & PPT (lin.) & Vicinity (para.) \\
  \hline
  2 & 4 & 9 & 6  \\
  3 & 77 & 81 & 60 \\
  4 & 244 & 256 & 211\\  
  5 & 613 & 625 & 547\\
  6 & 1272 & 1296 & 1166\\  
\end{tabular}
\end{ruledtabular}
\caption{Number of operators to be measured to estimate the expectation value of the entanglement witness in the parallel case for $D=2$ to $D=6$ for the parallel case and linear case using the PPT entanglement witness, and for the parallel case using the Vicinity entanglement witness.} \label{tab:table_2}

\end{table}

\indent A possible way to reduce the number of operators to be measured would be to find a sub-optimal witness with a lower number of terms to measure. One example is an approach similar to Bell inequalities. These do not detect all the entangled state as it is focused only on identifying states which cannot be explained through Local Hidden Variable (LHV) models. In \cite{Hyllus2005}, Hyllus et al. show that it is possible to convert an optimal entanglement witness into a CHSH type inequality. The resulting inequality detects non-LHV states optimally but not entangled LHV states.\\
\indent Therefore, there is a trade-off between the optimality of the entanglement witness (the finest entanglement witness being the optimal witness for a given entanglement detection problem) and the overall number of measurements required to test entanglement. The conversion method developed in Hyllus et al. only relates to qubits and does not take into account the impact of decoherence on the detectability of entanglement \cite{Hyllus2005}. However, for the qubit case, given the witness used only as three terms that need to be measured, CHSH types inequality are unlikely to provide a significant benefit.\footnote{Nguyen and Bernards provide a threshold value for the phase of the parallel set-up for which it would violate CHSH inequalities, including decoherence \cite{Nguyen2019}. Further research would be necessary in order to conduct a similar test for qudits.}  

\section{Time trade-off with decoherence} \label{sec:time_opt}

\begin{figure}
	\centering
		\includegraphics[width=1.0\columnwidth]{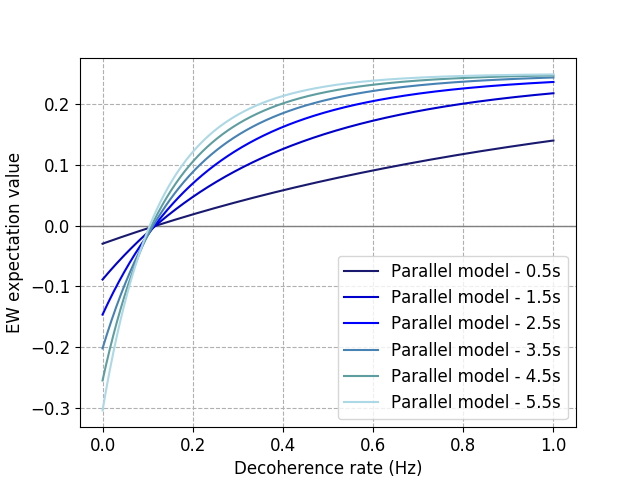}\par\medskip
	    \caption{Expectation value of PPT entanglement witness for different runtime of the experiment and as a function of the decoherence rate in the case $D=2$}\label{fig:deco_para_runtime2}
	\textit{}
	\end{figure}

\begin{figure}
	\centering
		\includegraphics[width=1.0\columnwidth]{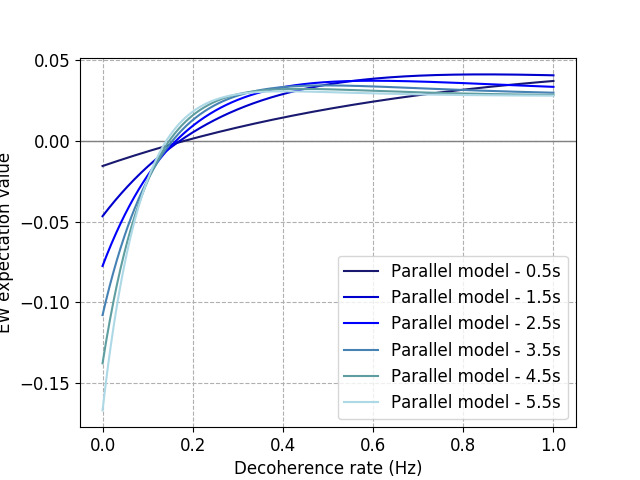}\par\medskip
		\caption{Expectation value of PPT entanglement witness for different runtime of the experiment and as a function of the decoherence rate in the case $D=6$}\label{fig:deco_para_runtime6}
	\textit{}
	\end{figure}

\indent In the original paper, Bose et al. suggest an experiment run-time of $\tau=2.5s$. In this appendix, We consider a simple way to estimate which run-time produce the highest resilience to decoherence (which run-time results in the expectation value of the entanglement witness reaching $0$ for the highest decoherence rate). Figure \ref{fig:deco_para_runtime2} presents the case of qubits, while Figure \ref{fig:deco_para_runtime6} present the case of 6-dimensional qudits. 

\indent On the qubit figure we can that additional time as little impact on how high the decoherence rate can be allowed to be. It does, however, offer more negative expectation values for the entanglement witness. If the experiment decoherence rate is estimated, one can then verify whether additional time can reduce the overall number of measurements required.

\indent The effect is more pronounced in the 6-dimensional qudits case as lower time offer negative expectation values of the witness for higher decoherence rate. 

\section{Computing confidence interval for a quantum observable} \label{sec:stats}

\indent In order to compute statistics related to the expectation value of a quantum observable, we first need to deconstruct the witness we are trying to estimate into a weighted sum of observable that can be directly measured, i.e. a set of Pauli strings. \\
\indent In particular, in the case of our two qudit system we have at most $D^4$ terms. These terms and the tensor product composite of the list of Gell-Mann matrices of dimension $D$ which are numbered $\mathcal{D} = D^2$. Any witness $\mathcal{W}$ can then be written as: 

\begin{equation}
    \mathcal{W} = \sum_i^\mathcal{D}\sum_j^\mathcal{D} c_{ij} \lambda_i^{(1)} \otimes \lambda_j^{(2)}
\end{equation}
\indent With $\lambda$ representing any Gell-Mann matrix (in the qubit case, these are the Pauli matrices: $\lambda \in \{\mathbb{I}, X, Y, Z \}$). Any tensor $w_{ij} = \lambda_i^{(1)} \otimes \lambda_j^{(2)}$ is a quantum observable that can be directly measured in an experiment. \\
\indent For a given number of measurements $M$, we can partially improve the overall variance of the observable by distributing these measurements in proportion to the weight of each term in the decomposition of the witness, such that, noting $c = \sum_i \sum_j \lvert c_{ij} \rvert$, we have: 
 \begin{align}
    M = \sum_i \sum_j M_{ij} \\
    M_{ij} =  \frac{\lvert c_{ij} \rvert}{c} M
\end{align}
\indent Considering that we conduct $M_{ij}$ measurements we can then determine the mean and variance of each of the term as follows: 
 \begin{equation}
    \overline{w_{ij}} = \sum_m^{M_{ij}} \frac{w_{ij}^{(m)}}{M_{ij}}
\end{equation}
 \begin{equation}
    \sigma_{ij}^2 = \sum_m^{M_{ij}} \frac{(\overline{w_{ij}} - w_{ij}^{(m)})^2}{(M_{ij} - 1)}
\end{equation}

\indent From there, we can compute the witness' mean and variance:
 \begin{equation}
    \overline{\mathcal{W}} = \sum_i^\mathcal{D} \sum_j^\mathcal{D} \sum_m^{M_{ij}} \frac{c_{ij} w_{ij}^{(m)}}{M_{ij}}
\end{equation}
 \begin{equation}
    \sigma_{\mathcal{W}}^2 = \sum_i^\mathcal{D} \sum_j^\mathcal{D} \sum_m^{M_{ij}} \lvert c_{ij}\rvert^2 \sigma_{ij}^2 
\end{equation}

\indent Finally, noting $\mathcal{M}$ the average number of measurements per term, we can compute the standard error of normally distributed measurement population as:

 \begin{equation}
    s_{\mathcal{W}} = \frac{\sigma_{\mathcal{W}}}{\sqrt{\mathcal{M}}}
\end{equation}

We then compute the confidence interval as: 
 \begin{equation}
    CI_\mathcal{W} = [\overline{\mathcal{W}} - \alpha s_{\mathcal{W}}, \overline{\mathcal{W}} + \alpha s_{\mathcal{W}}  ]
\end{equation}

\indent With $\alpha$ the t value corresponding to the desired level of confidence. \\
\indent For computation of the confidence level, we test against the null hypothesis $\mathcal{W} \ge \mu_0$, with $\mu_0 = 0$, and compute the t values following the traditional methods for one-sided t-test: 
 \begin{equation}
    t =\frac{\lvert \mathcal{W} - \mu_0 \rvert }{s_{\mathcal{W}}}
\end{equation}
\indent Confidence is then computed as $1 - p$, with $p$ the p-value corresponding to the t value obtained. \\

\section{Python model for the QGEM experiment} \label{sec:python}

\indent Our model for computation and modelling of the QGEM experiment is available on github under the name \href{https://github.com/JT76/Generalised_QGEM}{Generalised QGEM}

\end{appendices}

\end{document}